# Heavy Fermion metal $Fe_{16}N_2$ and its giant magnetic moment

Nian Ji[1,2], Xiaoqi Liu[1,3], Jian-Ping Wang[1,2,3*]

[1]The Center for Micromagnetics and Information Technologies (MINT), University of Minnesota, Minneapolis, Minnesota, 55455, USA
[2]Department of Physics, University of Minnesota, Minneapolis, Minnesota, 55455, USA
[3]Department of Electrical Engineer and Computer Science, University of Minnesota, Minneapolis, Minnesota, 55455, USA

**Abstract**

A new model is proposed for the strong ferromagnetism associated with partially localized orbitals in the $Fe_{16}N_2$ metallic system which draws substantially from models of heavy fermion metals. The basic idea is that the spatially isolated Fe-N clusters generate non-uniform charge density and increase the d-d electron interaction significantly, leading to a highly spin polarized configuration for low lying 3d orbitals. Simulation based on LDA+U method is performed to illustrate the correlation between enhanced U and giant magnetic moment.

**Context**

It is known that heavy fermion materials are always characterized by dual electronic states. They possess both localized orbitals which form the strong onsite magnetic moment and delocalized electrons which are responsible for the magnetic coupling and metallic properties. Traditional heavy fermion materials such as uranium containing compounds usually contain f electrons[1]. In some cases, transition metal oxide can also behave as heavy fermion metal. $LiV_2O_4$[2] is reported to be a 3d electron heavy fermion system, and this challenges the conventional understanding. Consequently, it is logical to explore electron behavior can occur in other materials. This research is further justified by the s-d model[3], which involves a mixture of localized and itinerant electrons and is used to describe ferromagnetic materials in general.

The iron nitrides discussed here were discovered many years ago. Interest in their magnetic properties originates from a discovery by Kim and Takahashi in the 1970's[4], when they reported a giant saturation magnetization (Ms) observed on the α" phase of $Fe_{16}N_2$. Almost 20 years later, research on this topic was stimulated by Sugita et al.[5] using single crystal single phase α"- $Fe_{16}N_2$ thin films grown epitaxially on GaAs or In doped GaAs by MBE. Saturation magnetization is found to be 3.2T, with a corresponding magnetic moment as high as 3.5$\mu_B$/Fe at low temperature. However, the subsequent investigations from many other groups on bulk and thin films samples containing α"- $Fe_{16}N_2$ led to conflicting results with Ms value less than 2.3T (2.5 $\mu_B$/Fe)[6]. Somewhat higher moments have been reported in films prepared by specially designed sputtering beams or facing target sputtering systems[7], reaching a value up to 2.9$\mu_B$/Fe and 3.0$\mu_B$/Fe, respectively. Due to such inconsistent and poorly reproducible reports, $Fe_{16}N_2$ has been regarded a mystery among magnetic researchers.

Initial theoretical models treated α"- $Fe_{16}N_2$ as a metal, and so used local spin density approximation (LSDA) based electronic structure calculations to predict its magnetic properties. These calculations do mot predict high magnetic moments in $Fe_{16}N_2$ even when first principle methods[8] or nonlocal correction have been considered. When nitrogen is added to iron in the LSDA scheme, the down spin electron changes distribution on different iron sites due to p-d hybridization between N and its nearest 6 iron neighbors[9]. Measurements on other iron nitrides phases suggest quantitative agreement with this down spin electron redistribution process, but it can not increase the average magnetic moment above 2.7 $\mu_B$/Fe for the α" phase. To rationalize the high moment reports, Lai et al.[10] postulated the existence of a strong correlation effect and obtained the magnetic moment under LDA+U method more closely resembling the high moment experimental data. A. Sakuma proposed a high moment scenario called the charge transfer model[11], which infers the existence of empty nitrogen orbitals near the Fermi level, which serve as charge hopping sites. This model predicts a high spin configuration of Fe and long range ferromagnetic order through an effective "double exchange" process. Lai and Sakuma's novel approaches provide further insights and require the α"- $Fe_{16}N_2$ system to have a dual electron configuration much like the heavy fermion system. Delocalized electrons provide metallic properties while localized electrons allow the application of Hund's rule, exchange interactions and direct Coulomb. Recently, our group has performed X-ray magnetic circular dichroism (XMCD) measurements on high-moment $Fe_{16}N_2$ samples and observed the localized 3d electron behavior[12]. In this paper, new model is proposed with partial localization of 3d electrons and strong ferromagnetism associated with non-uniform charge density distribution in the $Fe_{16}N_2$ system. By assuming that spatially isolated octahedral FeN clusters create non-uniform charge density distribution, the Hubbard U value consequently increases. As a result, the partial localization model can be used to estimate magnetic properties for $Fe_{16}N_2$ and is compatible with the LDA+U based calculation method. The effect of the d-d Coulomb interaction change will be used as an example to show the magnetic moment change.

The crystal structure of $Fe_{16}N_2$ is extensively discussed by K. H. Jack[13]. Roughly speaking, it contains both bcc- and fcc-like local Fe environment with some distortion. When N occupies interstitial sites of the bcc Fe lattice, a D4h symmetry Fe is formed by extending the octahedral coordination of Fe sites along the z axis, resulting in triple degenerate t2g orbital split into two well separated subsets (dxy and dxz+dyz). For example, bct Fe and the Fe closer to the N site in $Fe_4N$ possess such local symmetry. Before exploring the partial localization preliminary, LSDA calculations of the electronic structure of bct Fe were performed with the same lattice constant as that in $Fe_{16}N_2$. Fig.1 shows the partial DOS projected to the (dxz+dyz)/2 double degenerate orbital and dxy orbital. The width (W) of the overlapping area between the spin up and spin down DOS of the (dxz+dyz)/2 orbital is approximately 1.5ev and it represents equilibrium between kinetic energy cost and on site coulomb energy (U) cost. The overlapping region is even narrower in the dxy orbital (~1ev). In a real $Fe_{16}N_2$ system, two modifications will be considered. The first is the Fe-N hybridization effect that has been widely discussed in literature. In addition, the Fe sites with local D4h point group symmetry reduce to C4v and C1h symmetry due to the alternating occupancy of the N atoms. Unlike $Fe_4N$ or bct Fe, the corner Fe sites within those distorted Fe-N octahedrons lose the mirror reflection symmetry along certain directions, experiencing less symmetric crystal field due to the N coulomb charge. Therefore, an additional effect arises due to the symmetry break. The partial localization model focuses on such a reduced symmetry effect by assuming considerable charge density difference inside and outside the octahedral Fe-N cluster spherical region as schematically shown in fig.2. By assuming a uniform charge distribution within that sphere, the static potential can be obtained by solving Poison's equation as shown in the appendix. This potential may be written as:

$$\frac{na_0}{2r_0}Ry$$

with $n$ represents the total charge difference interior and exterior of the sphere, $a_0$ the Bohr radius, $Ry$ Rydberg constant, $r_0 = r_N + r_{Fe}$. This potential energy is created solely by the d-d electron interaction and adds to the on-site Coulomb interaction. The U values found in metallic Fe are typically 1ev. Assuming n=1 (~0.17 electron per Fe), the increase of the effective U value would be close to 1.5ev, which qualifies to the d-d electron interaction difference inside and outside the cluster. In analogy to the Stoner's ferromagnetism stability argument, such a high U/W ratio leads electron localization with a fully- polarized spin configuration for certain low lying orbitals such as dxy in within the Fe-N cluster. On the other hand, upper lying orbitals possess broad band width and less exchange splitting due to their spatial extension and strong hybridization with neighboring metallic Fe sites outside the sphere. This condition helps to maintain their band-like features that favor long-range ferromagnetic coupling and metallic behavior. The effect that non-uniform charge distribution indicates that $Fe_{16}N_2$ may be regarded as similar to the heavy Fermion system as reported by X. Liu et al.,[12]. To illustrate the partial localization model of $Fe_{16}N_2$, an electronic structure calculation was performed with the LDA+U method to compute magnetic moment. To simulate partial localization, the U value of Fe outside (Fe4d) the cluster sphere was fixed at 1 ev (typical value for metal) and the magnetic moment change was then plotted for three Fe sites as the U inside the cluster (same U is assumed for both 4e and 8h Fe) varied as shown in fig.3a. The J parameter was taken to be U/10. As the U increases, both 4e and 8h Fe are shown to favor rapid polarization while the 4d Fe site exhibits slow decay of the moment from 2.9 to 2.6μB/Fe and levels off at high U values. The vertical dashed line indicates a typical U value for simulations of metallic Fe and FeO in literature. An average magnetic moment of ~2.8μB/Fe can be obtained if a U=4ev is assumed. The effect of the d-d coulomb interaction on the electronic structure can be explained in the following. Plotted in Fig.3b is the partial DOS projected on dx2 orbital of Fe8h sites. As U begins to increase, the spin up band shifts downwards with more and more states occupied at the expense of all other orbitals in the process of self-consistency iteration. This orbital will eventually become fully occupied when U is sufficiently large. In that case, electrons on these orbitals exhibit nearly atomic configuration and the magnetic coupling mechanism is dominated by Hund's rule, which gives rise to a fully polarized spin configuration. It is also interesting to notice that the Fe4d sites reduce their magnetic moment as U increases. This suggests that the electron distribution on different iron sites changes significantly relative to the LSDA scheme, in which the high moment on Fe4d sites are due to the charge transfer from Fe4d to the Fe sites closer to the N sites.

The model proposed here has some similarities to charge transfers or double exchange mechanisms in the sense that N serves as hopping sites. However, it is seen that to achieve the magnetic moment value as high as experimentally claimed, it may require large U values, which could be criticized as unrealistic for a metal system. On-site coulomb interaction due to the screening of the itinerant band provides charge density difference through cluster isolation, but it is indeed difficult to estimate if the charge density difference due to the cluster isolation can provide such strong on-site coulomb interaction due to the screening of the itinerant band if deep level empty N orbital is regarded as hopping sites. Therefore, giant magnetic moment and localized d electron behavior may imply the existence of shallow localized electron states near Fermi level or perhaps defects states at the cluster region, either of which could cause increased charge density difference or reduce the total number of conducting electrons. The metallic behavior of the system still stands as long as the Fe 4d site remain itinerant. Also, these states should be related to the spatial distribution of Fe-N octahedral clusters that appear in Fe doped with low N concentration.

In conclusion, calculations using LDA+U method gives a theoretical justification for the proposed solution on the long standing $Fe_{16}N_2$ problem and additionally offers an explanation of the existence of the localized d electron discovery. The partial localization model provides insight to the additional effect brought about by N. In particular, isolated Fe-N octahedral clusters may induce non-uniform charge distribution and cause an increase of effective U value that generates both localized and itinerant d electrons. This feature allows the system to have global ferromagnetic coupling while locally maintain Hund's magnetic moments. Future work will be required to improve numerical results.

**Acknowledgement**

The work was partially supported by the U.S. Department of Energy, Office of Basic Energy Sciences under contract No. DE-AC02-98CH10886, National Science Foundation NNIN program at University of Minnesota. The authors would like to thank Prof. Jack Judy, Dr. Mark Kief and Dr. Yinjian Chen for the useful discussion.

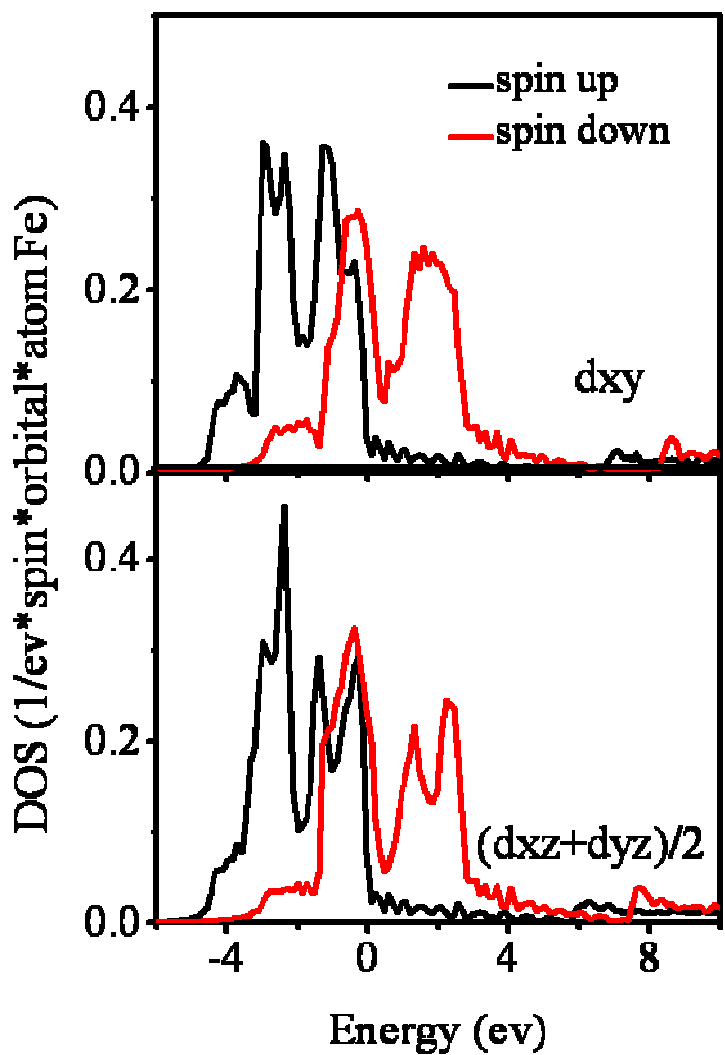

FIG.1. The density of states (DOS) of t2g band for bct Fe decomposed in dxy and (dxz+dyz)/2 obtained in the LSDA calculation

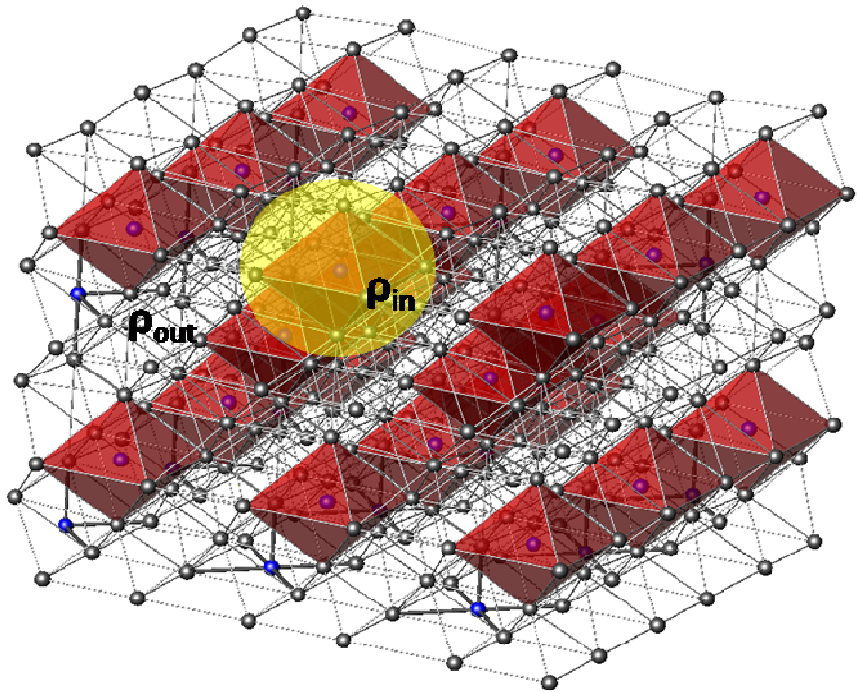

FIG.2. The crystal structure of $Fe_{16}N_2$. Fe-N octahedral clusters are highlighted in red color. The proposed charge density distribution is schematically marked by yellow sphere.

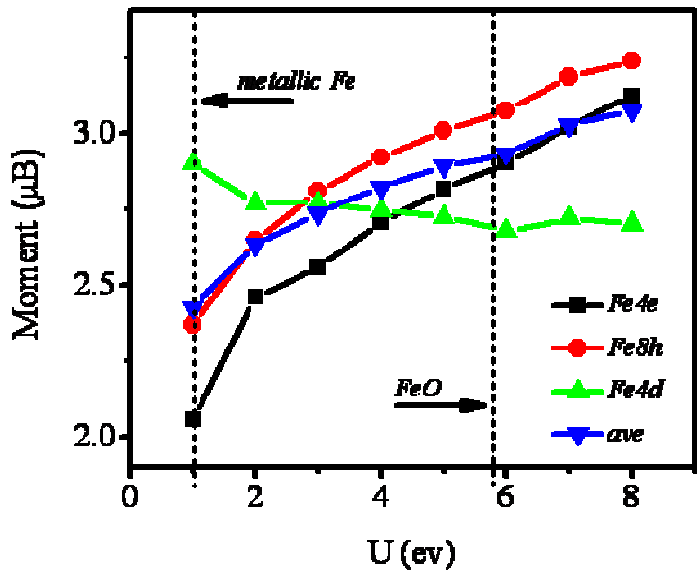

FIG. 3 a magnetic moment of different Fe site in $Fe_{16}N_2$ vs U as explained in text; b Partial density of states projected on $3dx2$ of Fe8h site in $Fe_{16}N_2$ for different U values.

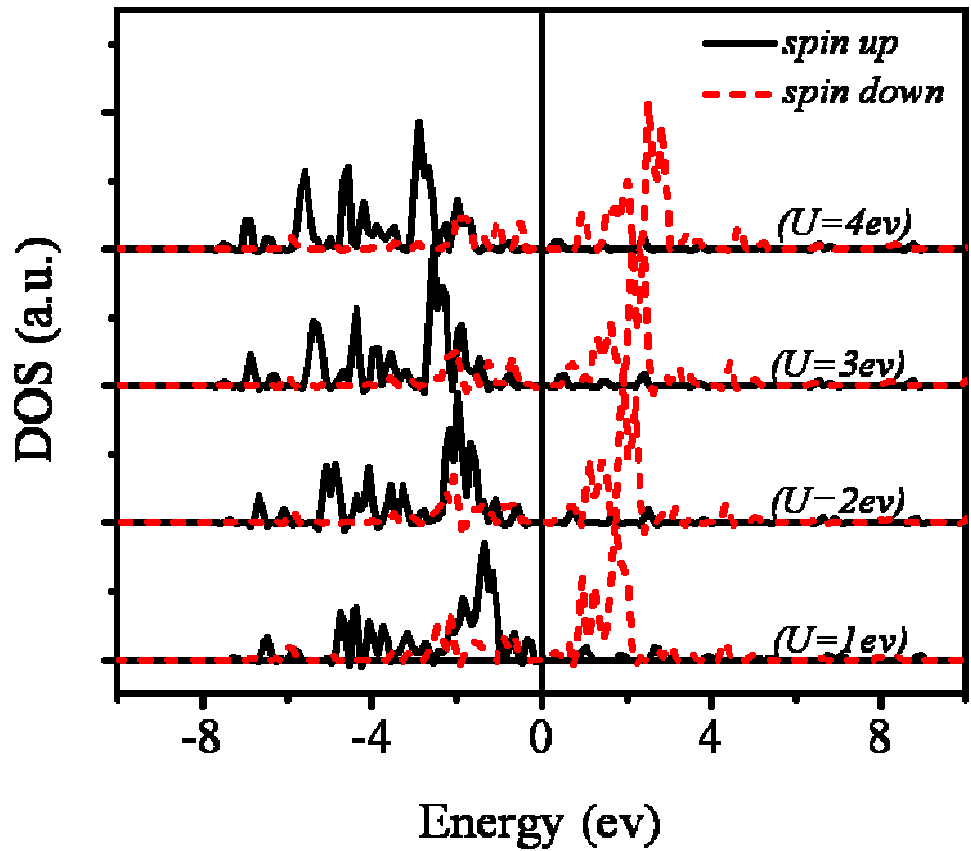

## Appendix

Poison equation relates the charge density and charge potential by:

$$\nabla^2 \varphi = \frac{\rho}{\varepsilon_0}$$

In spherical coordination and set boundary condition as

$$\varphi|_{r=r_0} = 0$$

$$\frac{\partial^2 \varphi}{\partial r^2} + \frac{2}{r}\frac{\partial \varphi}{\partial r} = -\frac{\rho}{\varepsilon_0}$$

One obtains

$$\varphi = \frac{\rho_{in}}{6\varepsilon_0}(r_0^2 - r^2)$$

$r_0$ is the radius of the cluster region approximately equals to $r_N + r_{Fe}$

The charge potential energy difference in and out the cluster should be roughly

$$E = e\varphi_{max} = \frac{e\rho_{in}}{6\varepsilon_0}r_0^2 = \frac{na_0}{2r_0}\frac{e^2}{4\pi\varepsilon_0 a_0} = \frac{na_0}{2r_0}Ry$$

n is the total number of d electrons involved within the cluster and Ry is Rydberg constant.